\documentclass[sigconf]{acmart}
\AtBeginDocument{%
  }

\usepackage{xcolor}
\usepackage{soul}
\usepackage{multirow}
\usepackage{tabularx}
\usepackage{booktabs}

\setcopyright{acmlicensed}
\copyrightyear{2018}
\acmYear{2025}
\acmDOI{XXXXXXX.XXXXXXX}
\acmConference[GLSVLSI '25]{Make sure to enter the correct
  conference title from your rights confirmation email}{June 30--July 05,
  2025}{New Orleans, LA, USA}
\acmISBN{978-1-4503-XXXX-X/2018/06}

\copyrightyear{2025}
\acmYear{2025}
\setcopyright{cc}
\setcctype{by}
\acmConference[GLSVLSI '25]{Great Lakes Symposium on VLSI 2025}{June 30-July 2, 2025}{New Orleans, LA, USA}
\acmBooktitle{Great Lakes Symposium on VLSI 2025 (GLSVLSI '25), June 30-July 2, 2025, New Orleans, LA, USA}
\acmDOI{10.1145/3716368.3735281}
\acmISBN{979-8-4007-1496-2/2025/06}

\begin{document}

\title{Transformers for Secure Hardware Systems: \\ Applications, Challenges, and Outlook}


\author{Banafsheh Saber Latibari}
\affiliation{%
  \institution{University of Arizona}
  \city{Tucson, AZ}
  \country{USA}}
\email{banafsheh@arizona.edu}

\author{Najmeh Nazari}
\affiliation{%
  \institution{University of California, Davis}
  \city{Davis, CA}
  \country{USA}}
  \email{nnazari@ucdavis.edu}

\author{Avesta Sasan}
\affiliation{%
 \institution{University of California, Davis}
 \city{Davis, CA}
 \country{USA}}
 \email{asasan@ucdavis.edu}

\author{Houman Homayoun}
\affiliation{%
 \institution{University of California, Davis}
 \city{Davis, CA}
 \country{USA}}
 \email{hhomayoun@ucdavis.edu}

\author{Pratik Satam}
\affiliation{%
  \institution{University of Arizona}
  \city{Tucson, AZ}
  \country{USA}}
  \email{pratiksatam@arizona.edu}

\author{Soheil Salehi}
\affiliation{%
  \institution{University of Arizona}
  \city{Tucson, AZ}
  \country{USA}}
\email{ssalehi@arizona.edu}

\author{Hossein Sayadi}
\affiliation{%
  \institution{California State University, Long Beach}
  \city{Long Beach, CA}
  \country{USA}}
\email{hossein.sayadi@csulb.edu}

\renewcommand{\shortauthors}{Saber Latibari et al.}

\begin{abstract}
 The rise of hardware-level security threats, such as side-channel attacks, hardware Trojans, and firmware vulnerabilities, demands advanced detection mechanisms that are more intelligent and adaptive. Traditional methods often fall short in addressing the complexity and evasiveness of modern attacks, driving increased interest in machine learning-based solutions. Among these, Transformer models, widely recognized for their success in natural language processing and computer vision, have gained traction in the security domain due to their ability to model complex dependencies, offering enhanced capabilities in identifying vulnerabilities, detecting anomalies, and reinforcing system integrity. 
 This survey provides a comprehensive review of recent advancements on the use of Transformers in hardware security, examining their application across key areas such as side-channel analysis, hardware Trojan detection, vulnerability classification, device fingerprinting, and firmware security. Furthermore, we discuss the practical challenges of applying Transformers to secure hardware systems, 
 and highlight opportunities and future research directions that position them as a foundation for next-generation hardware-assisted security. These insights pave the way for deeper integration of AI-driven techniques into hardware security frameworks, enabling more resilient and intelligent defenses.
\end{abstract}

\ccsdesc[500]{Security and privacy~Security in hardware}

\keywords{Hardware Systems, Transformer, Security, Threat Detection.}


\maketitle

\section{Introduction}
Hardware security has emerged as a critical pillar in the protection of modern computing systems, which are increasingly targeted by a diverse and evolving array of sophisticated threats \cite{nazari2024specscope}. 
Attacks such as side-channel exploits, hardware Trojans, and firmware-level vulnerabilities pose significant risks across a wide spectrum of platforms, from resource-constrained embedded devices to large-scale high-performance computing systems. 
These threats are particularly difficult to detect due to the complexity of modern hardware architectures and the ability of attackers to operate below the software stack \cite{nazari2024securing,wang2021enabling}. Traditional security mechanisms, such as rule-based heuristics, signature-based detection, and static analysis, often fail to generalize across evolving threat vectors and adapt to the dynamic nature of contemporary attacks. 

\begin{figure*}[h!]
    \centering
    \includegraphics[width=0.61\linewidth]{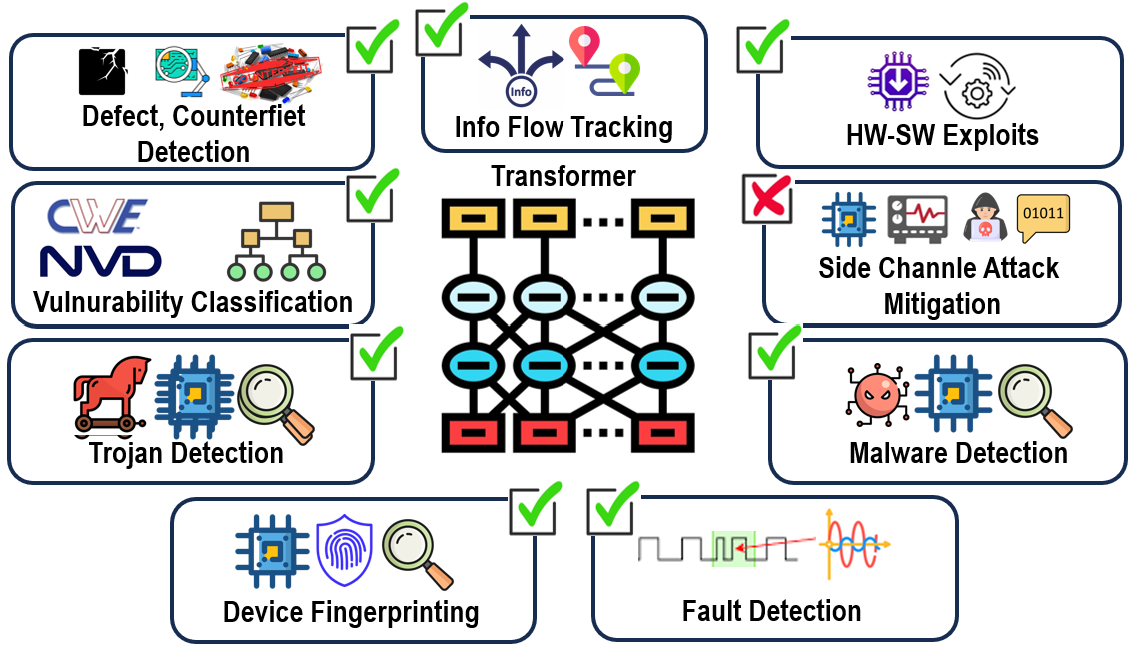}
    \caption{Scenarios where Transformers detect, classify, and mitigate vulnerabilities.}
    \label{security_fig}
\end{figure*}

As adversaries continue to develop more advanced and adaptive techniques, there is a pressing need for more intelligent, data-driven, and robust detection mechanisms. In recent years, to address these challenges, machine learning (ML)-based approaches have gained attention in the hardware security community. ML models can learn complex patterns from data, enabling anomaly detection, behavioral analysis, and predictive security measures \cite{sayadi2022towards}. 
However, conventional ML models often face limitations in this domain: they may struggle to capture long-range dependencies in hardware data, require extensive manual feature engineering, and frequently fail to scale effectively with the increasing heterogeneity and complexity of hardware systems data.

Transformer models have recently emerged as a powerful solution to these limitations. Originally designed for natural language processing, Transformers leverage self-attention mechanisms to capture intricate relationships within sequential data \cite{latibari2024transformers, saber2024iret, latibari2025optimizing}. This makes them especially well-suited for fast and accurate modeling of temporal and multi-modal patterns of the low-level data in hardware systems, without the need for high-overhead feature analysis. As depicted in Figure \ref{security_fig}, Transformers offer promising capabilities across a range of hardware security applications, from hardware Trojan detection and vulnerability assessment to malware detection and device fingerprinting. Their versatility and modeling power position them as a strong candidate for enabling next-generation, intelligent threat detection systems.

This paper presents a comprehensive survey of the application of Transformers in the context of hardware security. We review the current state of research, highlight key challenges and limitations, and outline future opportunities for advancing secure and intelligent hardware systems through this emerging paradigm. 
Section \ref{back} provides background on deep learning models relevant to this domain. Section \ref{security} reviews recent research applying Transformers to secure hardware systems. In Section  \ref{challenges}, we discuss the current challenges and opportunities, including practical constraints, architectural considerations, and potential directions. Lastly, Section \ref{con} concludes this study. 

\vspace{-1ex}
\section{Background on Transformers}\label{back}
Vaswani et al. \cite{vaswani2017attention} introduced the Transformer, which leverages the attention mechanism to model dependencies between input and output in machine translation. This approach eliminates the sequential nature of RNNs, enabling more efficient parallel processing and capturing long-range dependencies more effectively. Figure \ref{v_Transformer} illustrates the detailed computations within a Transformer architecture. 
This encoder-decoder-based architecture consists of multiple Transformer blocks, each containing a multi-head attention (MHA) module and a feed-forward network (FFN) module. Each block includes Layer Normalization (LayerNorm) and a residual connection. The MHA module first projects the input sequence using weight matrices $W_Q$, $W_K$, and $W_V$, generating query, key, and value representations. These representations are then split into $h$ heads, where each head has a hidden dimension of $d/h$, and processed as follows:
\begin{figure*}[h!]
    \centering
    \vspace{-1ex}
    \includegraphics[width=0.7\linewidth]{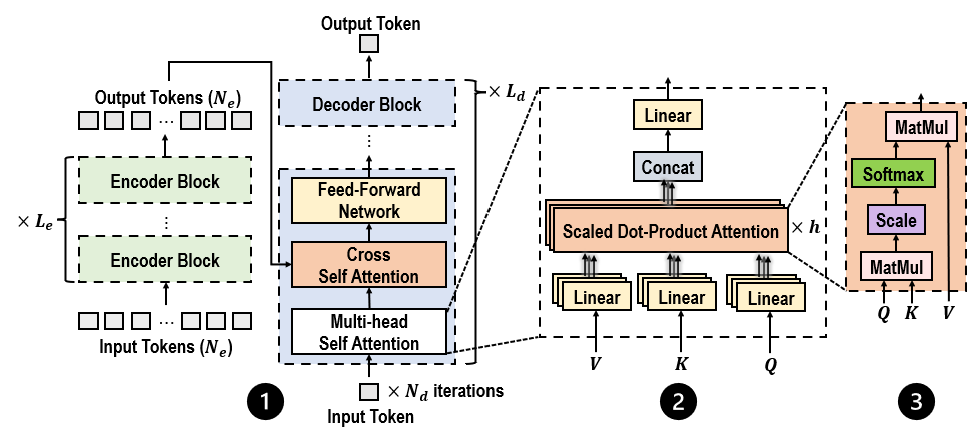}
    \caption{Architecture of the Vanilla Transformer, featuring (1) an Encoder-Decoder structure, (2) Multi-Head Self-Attention for capturing diverse contextual relationships, and (3) Scaled Dot-Product Attention for efficient information weighting.}
    \label{v_Transformer}
        \vspace{-1ex}
\end{figure*}

\begin{equation} \label{Linear}
\begin{split}
Q^{i} & =QW^{i}_{Q}\\
K^{i} & =KW^{i}_{K},  \hspace{8mm}  i \in heads\\      
V^{i} & =VW^{i}_{V}\\
\end{split}
\end{equation}

The scaled dot-product attention function computes the attention scores and output:

\begin{equation} \label{dot-att}
\centering
  O=SoftMax\left(\frac{QK^{T}}{\sqrt{d_{k}}}\right)V
\end{equation}

The attention outputs from all heads are concatenated along the hidden dimension $d$ and projected using $W_{out}$. The result is processed with LayerNorm and a residual connection to form the MHA output. The FFN module consists of two linear layers: the first projects the input from $d$ to a higher dimension, and the second projects it back to $d$.

\section{Transformers in Hardware Security} \label{security}

\subsection{Hardware Trojan Detection}
\textcolor{black}{Trojan is a malicious modification of a hardware design that can compromise functionality, integrity, or confidentiality. Since HTs are stealthy, ML-based models have been proposed to help with identification \cite{gubbi2023hardware}. }
For pre-silicon Hardware Trojan (HT) detection and localization, a Transformer-based method called HTrans is proposed. It utilizes a Graph Convolutional Network (GCN) in the preprocessing stage to ensure scalability across various design sizes. The model achieves 96.7\% F1 score for HT detection and 91.7\% accuracy for HT localization on the Trusthub benchmark, with detection completed in under a second at the Register Transfer Level (RTL) \cite{li2023htrans}. Authors in this paper propose a non-destructive, golden-chip-free transformer-based framework for Hardware Trojan Detection (HTD), utilizing Power Side-Channel (PSC) data. They apply Generative AI techniques such as GPT, BERT, and transformers to classify hardware trojans into Enabled, Disabled, and Triggered categories. The framework processes side-channel data through different transformer networks and achieves 87.74\% accuracy in HT detection, demonstrating superior performance compared to existing methods in identifying abnormal IC behaviors \cite{nasr2024improving}. 
TrojanFormer incorporates a unique message-passing scheme within a graph transformer network to enhance detection performance while reducing computational complexity. It achieves an average F1 score of 97.66\% on medium and small-scale datasets, surpassing other graph learning baseline models. On large-scale circuit datasets, TrojanFormer shows a 4\% performance improvement and an 18\% reduction in computational overhead, highlighting its effectiveness and efficiency in large-scale integrated circuit HT detection scenarios \cite{chen2024trojanformer}. 

TrojanWhisper \cite{faruque2024trojanwhisper} explores the potential of general-purpose LLMs for detecting HTs in Register Transfer Level (RTL) designs, including modules like SRAM, AES, and UART. The tool systematically evaluates state-of-the-art LLMs (GPT-4o, Gemini 1.5 pro, and Llama 3.1) in HT detection without prior fine-tuning. To address training data bias, perturbation techniques such as variable name obfuscation and design restructuring are implemented. The experimental results show perfect detection rates (100\% precision/recall) for GPT-4o and Gemini 1.5 pro, with performance degradation under code perturbation for all models, particularly in payload localization. 
This work demonstrates the potential of LLMs for hardware security applications \cite{faruque2024trojanwhisper}.

NtNDet leverages large-scale pre-trained NLP models. It introduces a method called Netlist-to-Natural-Language (NtN) to convert gate-level netlists into a format suitable for Natural Language Processing (NLP) models, applying the self-attention mechanism of Transformers to capture complex dependencies. Experiments on the Trust-Hub, TRIT-TC, and TRIT-TS benchmarks demonstrate that NtNDet outperforms existing methods, achieving improvements of 5.27\% in precision, 3.06\% in True Positive Rate (TPR), 0.01\% in True Negative Rate (TNR), and a 3.17\% increase in F1 score, setting a new state-of-the-art in HT detection \cite{kuang2025ntndet}. 
The work in \cite{chen2025lightweight} propose the TA-MobileViT lightweight model, which integrates the triplet attention (TA) mechanism with the MobileViT network to address the challenges in Hardware Trojan detection. This model enhances cross-channel interaction without increasing the parameter count, improving both classification accuracy and generalization ability. By combining convolutional and transformer blocks, TA-MobileViT effectively extracts both local and global features, achieving 100\% recognition accuracy for single Trojan types and 72.2\% accuracy for AES-600 detection. In multi-Trojan detection, the model reaches an impressive 97.04\% mean accuracy. Compared to other deep learning methods, TA-MobileViT offers better performance with fewer parameters, making it a highly competitive model for hardware Trojan detection.

\vspace{-7pt}

\subsection{Packaging Defect and Counterfeit Detection}
In the field of integrated circuit (IC) packaging and PCB design surface defect detection, ensuring high precision in identifying defects such as cracks, scratches, and contamination is crucial for maintaining product quality and manufacturing efficiency \cite{10643476, mohmadsalehi2022automated, mohamadsalehi2022expansion, chen2023comprehensive}.
The work in \cite{wang2024deep} addresses challenges in industrial IC surface defect detection, particularly the imbalance in information density due to data collection difficulties. The proposed hybrid model, SDDM, combines ResNet and Vision Transformer (ViT) to enhance defect detection by leveraging multi-channel image segmentation and convolution operations within patches. This approach improves the identification of high-information-density areas while optimizing computational efficiency for industrial applications. Experimental results show that SDDM achieves 98.6\% accuracy on imbalanced datasets, improving productivity in IC packaging and testing. 

Bhure et al. \cite{10792793} discuss the challenges posed by counterfeit components in the global semiconductor supply chain, which threaten product quality and reliability. The distributed nature of manufacturing and distribution increases the risk of fraud. To address this, they propose a Vision Transformer model for counterfeit IC detection, capable of classifying authentic samples and 11 types of counterfeit defects. Their model achieves 88\% accuracy on the test set, with attention map visualizations providing insights into its predictions \cite{10792793}.

\begin{table*}[h!] \small
\centering
\caption{System-Level Hardware Vulnerability Detection using Attention-Based Methods}
\vspace{-2ex}
\renewcommand{\arraystretch}{1.2}
\begin{tabularx}{\textwidth}{|l|l|X|X|}
\hline
\textbf{Category} & \textbf{Subsection} & \textbf{Description} & \textbf{Attention Usage} \\
\hline
\hline

\multirow{5}{*}{\textbf{Design-Time Security}} 
& Vulnerability Classification \cite{lin2023hw, fu2023vulexplainer} & Categorizing flaws in CWE and NVD datasets. & LLMs for processing datasets through a storytelling framework to suggest mitigations.  \\ \cline{2-4}
& Trojan Detection \cite{faruque2024trojanwhisper, li2023htrans} & Detect hardware trojans in 3P IPs or RTL. & Graph Attention Networks (GATs) to detect unusual paths. \\ \cline{2-4}
& Information Flow Tracking (IFT) \cite{mashnoor2025llm} & Trace propagation of sensitive data. & LLM-driven hierarchical
dependency analysis across modules. \\\cline{2-4}
& Defect \& Packaging Analysis \cite{wang2024deep, 10792793} & Detect packaging-related defects \& fake components. & Transformer models trained on SEM/X-ray images or electrical response sequences. \\
\hline

\multirow{6}{*}{\textbf{Runtime Security}} 
& Malware Detection \cite{seneviratne2022self, ravi2023vit4mal, deng2023transmalde} & Detect malicious behavior during operation. & Binaries converted into images, and ViTs are used instead. 
\\\cline{2-4}
& Side-channel Attack Detection \cite{hajra2024estranet, hajra2022transnet} & Identify information leakage via timing, EM, power, etc. & attention-modal on power traces, timing, etc. \\\cline{2-4}
& Fault Injection / Reliability-Aware Security \cite{shrivastwa2021multi} & Detect when injected faults cause vulnerabilities. & Multi-modal attention models combining error signals and system state. \\\cline{2-4}
& HW-SW Interface Exploits \cite{ali2025ddosvit, 9631481, YE2024103971} & Firmware and IoT systems face security risks. & Transformers model firmware, system calls, and network traffic as sequences to detect complex attacks. \\
\hline

\multirow{4}{*}{\textbf{Post-Deployment Assurance}} 
& Device Fingerprinting \cite{parpart2024transformer, shen2021radio} & Identify unique behavior profiles of devices. & Attention over power, Radio Frequency (RF) characteristics, or EM patterns. \\\cline{2-4}
& Remote Attestation \cite{10423646} & Verify system integrity over a network. & Transformers for cyber threat detection in IoT networks. \\\cline{2-4}
& Insider Threat Detection \cite{pal2023temporal} & Legitimate user abusing access privileges. & Attention models over system access logs and patterns. \\\cline{2-4}
\hline

\end{tabularx}
\end{table*}

\subsection{Side-Channel Attacks}
Recent advancements in leveraging Transformer models for the mitigation of side channel attacks (SCA) have significantly improved the robustness and effectiveness of cryptographic systems \cite{nazari2024architectural}. Berreby and Sauvage's work \cite{berreby2023investigating} employed the ANSSI Side-Channel Attack Database (ASCAD) and introduced a JAX-based framework specifically tailored for exploiting side-channel vulnerabilities in AES implementations. 

The SCAR framework \cite{srivastava2024scar} introduces a pre-silicon mitigation strategy utilizing Graph Neural Networks (GNNs) integrated with LLMs for early detection and automatic mitigation of power side-channel (PSC) leakage vulnerabilities at the Register-Transfer Level (RTL). SCAR converts RTL designs into control-data flow graphs (CDFGs), where nodes represent RTL basic blocks, and edges represent control flow. A GNN-based classification approach is applied for identifying modules susceptible to leakage, supported by LLM-generated mitigation code to automatically fortify identified vulnerable RTL code segments. This technique significantly enhances the early detection and mitigation of side-channel vulnerabilities, embedding security deeply within the hardware design cycle.
TransNet \cite{hajra2022transnet} offers a significant advance by incorporating shift invariance into transformer networks, a critical feature to effectively manage desynchronized power traces, and a common side channel countermeasure. TransNet leverages a modified Transformer architecture incorporating relative positional encoding and self-attention mechanisms that effectively capture dependencies among distant points of interest (PoIs). 

Building on TransNet, EstraNet \cite{hajra2024estranet} further optimizes the Transformer network specifically for SCA by achieving linear time and memory complexity. EstraNet introduces the GaussiP self-attention layer, an approach featuring relative positional encoding for enhanced shift-invariance and computational efficiency. 
Additionally, EstraNet employs a new layer-centering normalization method instead of traditional batch or layer normalization techniques, overcoming typical normalization challenges encountered in SCA applications. EstraNet demonstrated substantial robustness against masking, random delays, and clock jitter. Its scalability to very long traces (exceeding several thousand points) highlights its practical applicability in real-world cryptographic security scenarios.

\vspace{-5pt}

\subsection{Malware Detection}
Transformer models, with their powerful ability to capture intricate dependencies through self-attention mechanisms, have demonstrated significant potential in malware detection. 
A notable study by Seneviratne et al. introduces SHERLOCK \cite{seneviratne2022self}, a self-supervised deep learning framework that leverages ViT to detect Android malware. 
SHERLOCK adopts a self-supervised approach to learn robust feature representations from unlabeled binary samples. The binaries are converted into grayscale images, allowing the model to recognize malware patterns visually. 
To address the challenge of deploying effective malware detection on resource-constrained edge devices, Ravi et al. proposed ViT4Mal \cite{ravi2023vit4mal}. 
ViT4Mal also converts executable byte-code into images and employs a customized, lightweight ViT architecture designed explicitly for limited-resource hardware. 

Another innovative approach, TransMalDE, developed by Deng et al. \cite{deng2023transmalde}, targets the detection of IoT malware through a hierarchical Transformer-based framework. The TransMalDE framework migrates computationally intensive malware detection tasks from IoT devices to edge computing nodes, thus significantly reducing detection latency. 
TransMalDE captures the latent behavior patterns characteristic of evolving IoT malware by analyzing the textual semantic patterns that traditional methods might overlook.

Further extending Transformer models to utilize process resource utilization metrics, Natsos and Symeonidis present a dynamic malware detection technique \cite{natsos2025transformer}. 
The authors encode input data as sequences of processes, each represented by its resource metrics (CPU, memory, disk usage). 
Additionally, this research introduces dynamic malware signatures derived from resource metrics, revealing indirect malware activity indicators through cascading effects on system-wide processes.
These studies underscores the Transformer model's versatility and robustness in diverse malware detection contexts, from resource-rich cloud environments \cite{makrani2021security} to resource-constrained edge devices,  outperforming traditional ML techniques. 





\vspace{-5pt}

\subsection{Device Fingerprinting}
Similar to machine learning fingerprinting \cite{nazari2024llm}, device fingerprinting is also crucial for securing IoT communications against spoofing and cloning attacks by extracting unique radio frequency (RF) characteristics. Recent advancements in this domain extensively use deep learning methods for automatic feature extraction from transmitted signals, significantly improving the accuracy and robustness of device identification.
Wu et al. \cite{wu2018deep} leveraged an LSTM recurrent neural network for RF fingerprinting. Their model automatically captures intrinsic hardware-specific features, such as frequency drift and transient behaviors, achieving high accuracy even in environments with significant noise interference. Lee et al. \cite{9446687} further explored deep-learning-aided RF fingerprinting in Near Field Communication (NFC) systems and utilized neural networks, including CNN and RNN. 

Focusing on IoT scenarios, Jafari et al. \cite{8599826} used CNN and RNN to identify individual ZigBee devices. Their experiments across varied signal-to-noise ratio (SNR) conditions showed robust accuracy in distinguishing identical devices.
 Shen et al. \cite{shen2021radio} introduced Transformer-based models specifically for LoRa device fingerprinting, effectively managing variable-length signals and improving accuracy significantly through data augmentation and multi-packet inference, especially in low SNR conditions. 
Building upon Transformer architectures, Parpart et al. \cite{parpart2024transformer} proposed transformer masked autoencoders pre-trained on large-scale unlabeled RF data. Their method significantly enhanced classification accuracy, surpassing traditional CNN-based models, and demonstrated efficient handling of extensive datasets with thousands of devices. 






\subsection{Remote Attestation}
The rapid growth of IoT devices demands efficient mechanisms for detecting network-based attacks. SecurityBERT is a lightweight BERT-based model for cyber threat detection in IoT networks, using a novel Privacy-Preserving Fixed-Length Encoding (PPFLE) combined with a Byte-level tokenizer. Trained on the Edge-IIoTset dataset, it achieves 98.2\% accuracy across 14 attack types, outperforming traditional ML/DL methods while maintaining low inference time and small model size, making it ideal for resource-constrained IoT devices \cite{10423646}.

\subsection{Insider Threat Detection}
Insider threats pose a serious challenge for hardware systems, as they can exploit behavioral drift and data imbalance to stay hidden within normal activities. Detecting such threats is difficult due to the rarity of malicious behavior compared to regular user actions. \cite{pal2023temporal} proposes an insider threat detection approach using an ensemble of stacked-LSTM and stacked-GRU attention models trained on sequential activity logs. A new equally-weighted random sampling technique balances different threat categories, improving model fairness and performance. 

\subsection{Hardware-Software Inference Exploits}
The hardware-software interface is a possible source of security vulnerabilities caused by misconfiguration. Transformers can be applied to detect the vulnerability patterns through analyzing firmware routines, system calls, and instruction-level execution traces as sequences. The rise of IoT devices has further exposed firmware over-the-air (OTA) updates to threats like distributed denial of service (DDoS) attacks. Notably, the Mirai botnet exploited IoT vulnerabilities to launch massive attacks. Recent works, such as DDoSViT, leverage Vision Transformers (ViTs) by converting attack flows into images and training on datasets like CICIoT2023 and CICIoMT2024. These models achieved detection rates as high as 99.50\%, demonstrating the effectiveness of attention-based methods in identifying complex multi-vector attacks across the hardware-software boundary \cite{ali2025ddosvit}. FirmVulSeeker is a firmware vulnerability search tool that leverages BERT pretraining and a Siamese network to match semantically similar functions \cite{9631481}. SLFHunter is a LLM-based framework for detecting command injection vulnerabilities in embedded Linux firmware. By identifying sensitive dynamically linked library functions (DLLFs) with ChatGPT, it marks new sinks for static analysis tools like EmTaint \cite{YE2024103971}.

\subsection{Information Flow Tracking}
Information flow tracking (IFT) is a method for monitoring how data moves through a system. It aims to detect unauthorized or suspicious flows that may lead to vulnerabilities such as data leakage or privilege escalation. LLM-IFT leverages LLM-driven hierarchical dependency analysis across intra- and inter-module levels to overcome the scalability and adaptability limitations of traditional IFT methods. The approach achieves 100\% success in confidentiality and integrity checks on Trust-Hub benchmarks, shows the effectiveness of LLMs for security analysis in integrated circuits \cite{mashnoor2025llm}.

\subsection{Fault Detection}
Fault attacks, such as electromagnetic (EM) injection, voltage glitching, and clock manipulation, are powerful techniques for inducing errors in hardware and bypassing security mechanisms. Smart Monitor is a hardware framework that uses on-chip sensors and an AI core to detect and classify fault attacks like electromagnetic (EM) and clock-glitch (CG) injections. It achieves 92\% detection accuracy and 78\% classification accuracy with zero false positives \cite{shrivastwa2021multi}. 

\subsection{Vulnerability Classification}
With the increasing number of reported vulnerabilities, datasets like the National Vulnerability Database (NVD), Common Weakness Enumeration (CWE), and Common Vulnerabilities and Exposures (CVE) have become essential for developing effective security solutions. However, extracting meaningful insights from these vast datasets is not easy, making it necessary to apply AI techniques for effective vulnerability classification.
VulExplainer is a deep learning approach for classifying and explaining vulnerabilities using a Transformer-based hierarchical distillation framework. It improves classification accuracy by 5\%–29\% and is compatible with models like CodeBERT, GraphCodeBERT, and CodeGPT without architectural modifications \cite{fu2023vulexplainer}.
The HW-V2W-Map Framework is a machine learning-based approach for mapping hardware vulnerabilities to weaknesses, with a focus on IoT security. It incorporates an Ontology-driven Storytelling framework to track vulnerability trends. Additionally, it leverages GPT-based LLMs to generate mitigation strategies, helping to predict and prevent future vulnerabilities \cite{lin2023hw}. To enhance Industrial Control System (ICS) security, this study proposes deep learning-based automated vulnerability categorization. Given the limitations of national NVDs, the authors demonstrate that LSTM-tuned BERT models achieve superior precision, F1 score, accuracy, and recall. This approach strengthens cyber threat intelligence (CTI) and improves attack mitigation in ICS environments \cite{marali2024hybrid}.

\section{Challenges and Opportunities}\label{challenges}

\subsection{Computational Cost and Overhead}

Transformer models offer strong performance for hardware security tasks like side-channel analysis, Trojan detection, and malware classification, but they come with high computational costs. Their self-attention mechanism has quadratic complexity, making it difficult to deploy them directly on resource-constrained systems like embedded devices or FPGAs. To address this, some models like EstraNet and TA-MobileViT use efficient attention mechanisms or lightweight architectures that reduce memory and power usage while maintaining detection accuracy.
Various optimization techniques help reduce overhead. These include pruning unnecessary weights, quantizing models to use lower precision numbers, and using efficient attention mechanisms like sparse or linear attention. Such changes can make models faster and more suitable for low-power environments with minimal impact on performance. Designs often combine convolution layers with attention or reduce input size early to ease processing.

In real-world deployments, heavy computation is often offloaded to edge servers to lighten the burden on local devices, as seen in frameworks like TransMalDE. While this reduces latency and energy use on-device, it can introduce new trade-offs like network dependency. Overall, by applying the right balance of architectural changes and deployment strategies, Transformers can be effectively used for secure, real-time applications in hardware systems.

\vspace{-1ex}
\subsection{Lack of Explainability}
One major challenge in using transformers for hardware vulnerability detection and mitigation is the lack of explainability and interpretability. These models function as black boxes, making it difficult to understand their decision-making process, which is crucial for security-critical applications. Without clear insights into how vulnerabilities are identified, debugging and trust in the system become significant concerns. Additionally, the absence of interpretability hinders the ability to validate predictions and ensure robustness against adversarial attacks. Solutions to address these limitations include attention visualization, feature attribution methods, saliency-guided training to enhance model focus on relevant features, and hybrid approaches that integrate symbolic reasoning for improved transparency \cite{karkehabadi2024hlgm}.

\subsection{Vulnerability Against Adversarial Attacks}
Despite their effectiveness, deep learning models remain highly susceptible to adversarial attacks, and attention-based architectures, including Transformers, are no exception. Adversarial attacks involve crafting subtly perturbed inputs that mislead the model into making incorrect predictions or failing to detect malicious behavior \cite{mirzaeian2022adaptive}. This vulnerability poses a significant challenge for security-critical applications such as malware detection, where adversaries can potentially exploit model weaknesses to evade detection \cite{dinakarrao2019adversarial}.

To address these risks, researchers have explored defense mechanisms such as adversarial training, which improves robustness by incorporating adversarial examples during model training. He et al. \cite{he-DAC2024beyond} has demonstrated that ML-based malware detection systems are susceptible to adversarial attacks, especially those operating on structured tabular data like processor performance counters. To address this, a multi-phased defense framework based on Deep Reinforcement Learning (DRL) was introduced, combining adversarial training with real-time attack pattern prediction and dynamic defense assignment via a UCB-guided controller, This approach significantly improved detection robustness, achieving up to an 86\% increase in F1-score. 

In addition, recent studies have also highlighted the vulnerability of Transformer models to Bit-Flip Attacks (BFAs), where adversaries manipulate a small number of model parameters at the binary level to degrade performance. A novel defense strategy, Forget and Rewire (FaR) \cite{nazari2024forget}, introduces targeted rewiring in Transformer Linear layers to obscure critical neurons and redistribute computation, significantly improving model robustness with minimal accuracy loss.  
Additionally, hybrid approaches that combine deep learning with rule-based or statistical methods may offer enhanced reliability by introducing complementary detection layers. As Transformer-based models are increasingly adopted in hardware and system security, developing robust, interpretable, and attack-resilient architectures remains a key research priority.


\subsection{Availability of Dataset}
One of the main challenges in applying attention-based models is their reliance on large-scale, high-quality datasets for effective training. However, in this research domain, such datasets are often scarce or inaccessible. In many cases, access to design files or detailed system information needed to construct meaningful datasets is restricted, as hardware companies are unwilling to share proprietary or sensitive data. This lack of transparency significantly hinders dataset creation and model development. To address this limitation, researchers must explore alternative solutions such as synthetic data generation, simulation-based environments, or data augmentation techniques to produce diverse and representative training samples.

\vspace{-1ex}
\subsection{Real-Time Hardware-Level Threat Detection}
Hardware-level threat detection has gained momentum as a robust complement to traditional software-based defenses, offering deeper visibility into runtime behavior through low-level microarchitectural signals. Hardware-Assisted Malware Detection (HMD) techniques leverage sources such as Hardware Performance Counters (HPCs) and other on-chip telemetry to collect fine-grained execution traces, enabling the identification of anomalous and potentially malicious activities in real-time \cite{demme2013feasibility,sayadi-DATE2019-2smart,sayadi-ensemble-DAC18,gao2021adaptive}. 
Transformers present compelling opportunities for advancing HMDs. Their self-attention mechanism enables context-aware modeling of execution behavior, allowing the capture of subtle correlations in tabular hardware data that may be overlooked by traditional models. This capability is particularly valuable for detecting complex and stealthy attack patterns, including zero-day malware that manifests across temporal sequences or diverse microarchitectural events \cite{sayadi2020stealthminer,he2024guarding}. Moreover, Transformers’ ability to handle multi-modal inputs makes them well-suited for fusing heterogeneous side-channel signals (e.g., HPCs, power telemetry) into a unified representation, enabling a more in-depth analysis of system behavior under attack.

However, there exist several challenges in employing Transformers for hardware-assisted malware detection. Transformers are computationally demanding, limiting their deployment in resource-constrained environments where HMD is most needed. Their use with structured tabular data, such as HPC traces, also requires careful architectural tuning and feature encoding. Additionally, interpretability and robustness remain critical concerns, especially in security-sensitive applications where transparency is essential \cite{pan2022-explainHMD,sayadi-ISQED2024-survey}. 
Despite these challenges, the continued evolution of Tiny Transformers, efficient attention mechanisms, and self-supervised pretraining on hardware execution traces presents a path forward. Future work could explore hybrid architectures that balance accuracy and efficiency, domain-specific pretraining to improve generalization, and adversarial training techniques to strengthen resilience. As malware grows more sophisticated, the integration of Transformers into HMD systems, and more broadly across hardware-assisted security domains, holds the potential to unlock next-generation, adaptive, and intelligent threat detection frameworks.

\vspace{-1ex}
\section{Concluding Remarks}\label{con}
This paper presents a comprehensive review of recent developments in the application of Transformer-based models for hardware security. We examine their application across a wide range of critical domains, including side-channel analysis, hardware Trojan detection, device fingerprinting, vulnerability classification, malware detection, and firmware security. With their ability to capture complex dependencies and process multi-modal inputs, Transformers represent a significant shift in the design of intelligent, context-aware threat detection mechanisms at the hardware level. Despite their considerable promise, the application of Transformers to hardware-based security remains an emerging area with several open challenges, each of which also presents opportunities for future exploration. Key issues include computational overhead, susceptibility to adversarial attacks, limited interpretability, limited publicly available datasets, and the need for enhanced adaptability to structured hardware telemetry. Addressing these limitations will be critical to enabling scalable, trustworthy, and widely deployable Transformer-based security solutions, and offers a rich landscape for advancing research in secure and intelligent hardware systems.

\begin{acks}
This work is supported in part by the National Science Foundation under Award No. 2139034.
\end{acks}

\bibliographystyle{ACM-Reference-Format}
\bibliography{main}


\begin{thebibliography}{60}


\ifx \showCODEN    \undefined \def \showCODEN     #1{\unskip}     \fi
\ifx \showISBNx    \undefined \def \showISBNx     #1{\unskip}     \fi
\ifx \showISBNxiii \undefined \def \showISBNxiii  #1{\unskip}     \fi
\ifx \showISSN     \undefined \def \showISSN      #1{\unskip}     \fi
\ifx \showLCCN     \undefined \def \showLCCN      #1{\unskip}     \fi
\ifx \shownote     \undefined \def \shownote      #1{#1}          \fi
\ifx \showarticletitle \undefined \def \showarticletitle #1{#1}   \fi
\ifx \showURL      \undefined \def \showURL       {\relax}        \fi
\providecommand\bibfield[2]{#2}
\providecommand\bibinfo[2]{#2}
\providecommand\natexlab[1]{#1}
\providecommand\showeprint[2][]{arXiv:#2}

\bibitem[Ali et~al\mbox{.}(2025)]%
        {ali2025ddosvit}
\bibfield{author}{\bibinfo{person}{Muhammad Ali}, \bibinfo{person}{Yasir Saleem}, \bibinfo{person}{Sadaf Hina}, {and} \bibinfo{person}{Ghalib~A Shah}.} \bibinfo{year}{2025}\natexlab{}.
\newblock \showarticletitle{DDoSViT: IoT DDoS attack detection for fortifying firmware Over-The-Air (OTA) updates using vision transformer}.
\newblock \bibinfo{journal}{\emph{Internet of Things}} (\bibinfo{year}{2025}), \bibinfo{pages}{101527}.
\newblock


\bibitem[Berreby and Sauvage(2023)]%
        {berreby2023investigating}
\bibfield{author}{\bibinfo{person}{Yoha{\"\i}-Eliel Berreby} {and} \bibinfo{person}{Laurent Sauvage}.} \bibinfo{year}{2023}\natexlab{}.
\newblock \showarticletitle{Investigating efficient deep learning architectures for side-channel attacks on AES}.
\newblock \bibinfo{journal}{\emph{arXiv preprint arXiv:2309.13170}} (\bibinfo{year}{2023}).
\newblock


\bibitem[Bhure et~al\mbox{.}(2024)]%
        {10792793}
\bibfield{author}{\bibinfo{person}{Chaitanya Bhure}, \bibinfo{person}{Dhruvakumar Aklekar}, \bibinfo{person}{Wenjie Che}, {and} \bibinfo{person}{Fareena Saqib}.} \bibinfo{year}{2024}\natexlab{}.
\newblock \showarticletitle{Vision Transformers for Counterfeit IC Detection}. In \bibinfo{booktitle}{\emph{2024 IEEE Physical Assurance and Inspection of Electronics (PAINE)}}. \bibinfo{pages}{1--7}.
\newblock
\href{https://doi.org/10.1109/PAINE62042.2024.10792793}{doi:\nolinkurl{10.1109/PAINE62042.2024.10792793}}


\bibitem[Chen et~al\mbox{.}(2024)]%
        {chen2024trojanformer}
\bibfield{author}{\bibinfo{person}{Menghui Chen}, \bibinfo{person}{Xiaoyong Kou}, {and} \bibinfo{person}{Gongxuan Zhang}.} \bibinfo{year}{2024}\natexlab{}.
\newblock \showarticletitle{TrojanFormer: Resource-Efficient Hardware Trojan Detection Using Graph Transformer Network}. In \bibinfo{booktitle}{\emph{2024 7th International Conference on Electronics Technology (ICET)}}. IEEE, \bibinfo{pages}{165--170}.
\newblock


\bibitem[Chen et~al\mbox{.}(2025)]%
        {chen2025lightweight}
\bibfield{author}{\bibinfo{person}{Shouhong Chen}, \bibinfo{person}{Guanxiang Qin}, \bibinfo{person}{Ying Lu}, \bibinfo{person}{Tao Wang}, {and} \bibinfo{person}{Xingna Hou}.} \bibinfo{year}{2025}\natexlab{}.
\newblock \showarticletitle{A lightweight Hardware Trojan detection approach in the waveform diagram based on MobileViT and attention mechanism}.
\newblock \bibinfo{journal}{\emph{The Journal of Supercomputing}}  \bibinfo{volume}{81} (\bibinfo{year}{2025}), \bibinfo{pages}{580}.
\newblock


\bibitem[Chen et~al\mbox{.}(2023)]%
        {chen2023comprehensive}
\bibfield{author}{\bibinfo{person}{Xing Chen}, \bibinfo{person}{Yonglei Wu}, \bibinfo{person}{Xingyou He}, {and} \bibinfo{person}{Wuyi Ming}.} \bibinfo{year}{2023}\natexlab{}.
\newblock \showarticletitle{A comprehensive review of deep learning-based PCB defect detection}.
\newblock \bibinfo{journal}{\emph{IEEE Access}}  \bibinfo{volume}{11} (\bibinfo{year}{2023}), \bibinfo{pages}{139017--139038}.
\newblock


\bibitem[Demme et~al\mbox{.}(2013)]%
        {demme2013feasibility}
\bibfield{author}{\bibinfo{person}{John Demme}, \bibinfo{person}{Matthew Maycock}, \bibinfo{person}{Jared Schmitz}, \bibinfo{person}{Adrian Tang}, \bibinfo{person}{Adam Waksman}, \bibinfo{person}{Simha Sethumadhavan}, {and} \bibinfo{person}{Salvatore Stolfo}.} \bibinfo{year}{2013}\natexlab{}.
\newblock \showarticletitle{On the feasibility of online malware detection with performance counters}.
\newblock \bibinfo{journal}{\emph{ACM SIGARCH computer architecture news}} \bibinfo{volume}{41}, \bibinfo{number}{3} (\bibinfo{year}{2013}), \bibinfo{pages}{559--570}.
\newblock


\bibitem[Deng et~al\mbox{.}(2023)]%
        {deng2023transmalde}
\bibfield{author}{\bibinfo{person}{Xiaoheng Deng}, \bibinfo{person}{Zhe Wang}, \bibinfo{person}{Xinjun Pei}, {and} \bibinfo{person}{Kaiping Xue}.} \bibinfo{year}{2023}\natexlab{}.
\newblock \showarticletitle{TransMalDE: an effective transformer based hierarchical framework for IoT malware detection}.
\newblock \bibinfo{journal}{\emph{IEEE Transactions on Network Science and Engineering}} \bibinfo{volume}{11}, \bibinfo{number}{1} (\bibinfo{year}{2023}), \bibinfo{pages}{140--151}.
\newblock


\bibitem[Dinakarrao et~al\mbox{.}(2019)]%
        {dinakarrao2019adversarial}
\bibfield{author}{\bibinfo{person}{Sai Manoj~Pudukotai Dinakarrao} {et~al\mbox{.}}} \bibinfo{year}{2019}\natexlab{}.
\newblock \showarticletitle{Adversarial attack on microarchitectural events based malware detectors}. In \bibinfo{booktitle}{\emph{Proceedings of the 56th Annual Design Automation Conference 2019}}. \bibinfo{pages}{1--6}.
\newblock


\bibitem[Faruque et~al\mbox{.}(2024)]%
        {faruque2024trojanwhisper}
\bibfield{author}{\bibinfo{person}{Md~Omar Faruque}, \bibinfo{person}{Peter Jamieson}, \bibinfo{person}{Ahmad Patooghy}, {and} \bibinfo{person}{Abdel-Hameed~A Badawy}.} \bibinfo{year}{2024}\natexlab{}.
\newblock \showarticletitle{TrojanWhisper: Evaluating Pre-trained LLMs to Detect and Localize Hardware Trojans}.
\newblock \bibinfo{journal}{\emph{arXiv preprint arXiv:2412.07636}} (\bibinfo{year}{2024}).
\newblock


\bibitem[Ferrag et~al\mbox{.}(2024)]%
        {10423646}
\bibfield{author}{\bibinfo{person}{Mohamed~Amine Ferrag} {et~al\mbox{.}}} \bibinfo{year}{2024}\natexlab{}.
\newblock \showarticletitle{Revolutionizing Cyber Threat Detection With Large Language Models: A Privacy-Preserving BERT-Based Lightweight Model for IoT/IIoT Devices}.
\newblock \bibinfo{journal}{\emph{IEEE Access}}  \bibinfo{volume}{12} (\bibinfo{year}{2024}), \bibinfo{pages}{23733--23750}.
\newblock
\href{https://doi.org/10.1109/ACCESS.2024.3363469}{doi:\nolinkurl{10.1109/ACCESS.2024.3363469}}


\bibitem[Fu et~al\mbox{.}(2023)]%
        {fu2023vulexplainer}
\bibfield{author}{\bibinfo{person}{Michael Fu}, \bibinfo{person}{Van Nguyen}, \bibinfo{person}{Chakkrit~Kla Tantithamthavorn}, \bibinfo{person}{Trung Le}, {and} \bibinfo{person}{Dinh Phung}.} \bibinfo{year}{2023}\natexlab{}.
\newblock \showarticletitle{Vulexplainer: A transformer-based hierarchical distillation for explaining vulnerability types}.
\newblock \bibinfo{journal}{\emph{IEEE Transactions on Software Engineering}} \bibinfo{volume}{49}, \bibinfo{number}{10} (\bibinfo{year}{2023}), \bibinfo{pages}{4550--4565}.
\newblock


\bibitem[Gao et~al\mbox{.}(2021)]%
        {gao2021adaptive}
\bibfield{author}{\bibinfo{person}{Yifeng Gao}, \bibinfo{person}{Hosein~Mohammadi Makrani}, \bibinfo{person}{Mehrdad Aliasgari}, \bibinfo{person}{Amin Rezaei}, \bibinfo{person}{Jessica Lin}, \bibinfo{person}{Houman Homayoun}, {and} \bibinfo{person}{Hossein Sayadi}.} \bibinfo{year}{2021}\natexlab{}.
\newblock \showarticletitle{Adaptive-hmd: Accurate and cost-efficient machine learning-driven malware detection using microarchitectural events}. In \bibinfo{booktitle}{\emph{2021 IEEE 27th International Symposium on On-Line Testing and Robust System Design (IOLTS)}}. IEEE, \bibinfo{pages}{1--7}.
\newblock


\bibitem[Gubbi et~al\mbox{.}(2023)]%
        {gubbi2023hardware}
\bibfield{author}{\bibinfo{person}{Kevin~Immanuel Gubbi}, \bibinfo{person}{Banafsheh Saber~Latibari}, \bibinfo{person}{Anirudh Srikanth}, \bibinfo{person}{Tyler Sheaves}, \bibinfo{person}{Sayed~Arash Beheshti-Shirazi}, \bibinfo{person}{Sai~Manoj PD}, \bibinfo{person}{Satareh Rafatirad}, \bibinfo{person}{Avesta Sasan}, \bibinfo{person}{Houman Homayoun}, {and} \bibinfo{person}{Soheil Salehi}.} \bibinfo{year}{2023}\natexlab{}.
\newblock \showarticletitle{Hardware trojan detection using machine learning: A tutorial}.
\newblock \bibinfo{journal}{\emph{ACM Transactions on Embedded Computing Systems}} \bibinfo{volume}{22}, \bibinfo{number}{3} (\bibinfo{year}{2023}), \bibinfo{pages}{1--26}.
\newblock


\bibitem[Hajra et~al\mbox{.}(2024)]%
        {hajra2024estranet}
\bibfield{author}{\bibinfo{person}{Suvadeep Hajra} {et~al\mbox{.}}} \bibinfo{year}{2024}\natexlab{}.
\newblock \showarticletitle{Estranet: An efficient shift-invariant transformer network for side-channel analysis}.
\newblock \bibinfo{journal}{\emph{IACR Transactions on Cryptographic Hardware and Embedded Systems}} \bibinfo{volume}{2024}, \bibinfo{number}{1} (\bibinfo{year}{2024}), \bibinfo{pages}{336--374}.
\newblock


\bibitem[Hajra et~al\mbox{.}(2022)]%
        {hajra2022transnet}
\bibfield{author}{\bibinfo{person}{Suvadeep Hajra}, \bibinfo{person}{Sayandeep Saha}, \bibinfo{person}{Manaar Alam}, {and} \bibinfo{person}{Debdeep Mukhopadhyay}.} \bibinfo{year}{2022}\natexlab{}.
\newblock \showarticletitle{Transnet: Shift invariant transformer network for side channel analysis}. In \bibinfo{booktitle}{\emph{International Conference on Cryptology in Africa}}. Springer, \bibinfo{pages}{371--396}.
\newblock


\bibitem[He et~al\mbox{.}(2024a)]%
        {he-DAC2024beyond}
\bibfield{author}{\bibinfo{person}{Zhangying He}, \bibinfo{person}{Houman Homayoun}, {and} \bibinfo{person}{Hossein Sayadi}.} \bibinfo{year}{2024}\natexlab{a}.
\newblock \showarticletitle{Beyond conventional defenses: Proactive and adversarial-resilient hardware malware detection using deep reinforcement learning}. In \bibinfo{booktitle}{\emph{Proceedings of the 61st ACM/IEEE Design Automation Conference}}. \bibinfo{pages}{1--6}.
\newblock


\bibitem[He et~al\mbox{.}(2024b)]%
        {he2024guarding}
\bibfield{author}{\bibinfo{person}{Zhangying He}, \bibinfo{person}{Houman Homayoun}, {and} \bibinfo{person}{Hossein Sayadi}.} \bibinfo{year}{2024}\natexlab{b}.
\newblock \showarticletitle{Guarding against the unknown: Deep transfer learning for hardware image-based malware detection}.
\newblock \bibinfo{journal}{\emph{Journal of Hardware and Systems Security}} \bibinfo{volume}{8}, \bibinfo{number}{2} (\bibinfo{year}{2024}), \bibinfo{pages}{61--78}.
\newblock


\bibitem[Jafari et~al\mbox{.}(2018)]%
        {8599826}
\bibfield{author}{\bibinfo{person}{Hossein Jafari}, \bibinfo{person}{Oluwaseyi Omotere}, \bibinfo{person}{Damilola Adesina}, \bibinfo{person}{Hsiang-Huang Wu}, {and} \bibinfo{person}{Lijun Qian}.} \bibinfo{year}{2018}\natexlab{}.
\newblock \showarticletitle{IoT Devices Fingerprinting Using Deep Learning}. In \bibinfo{booktitle}{\emph{MILCOM 2018 - 2018 IEEE Military Communications Conference (MILCOM)}}. \bibinfo{pages}{1--9}.
\newblock
\href{https://doi.org/10.1109/MILCOM.2018.8599826}{doi:\nolinkurl{10.1109/MILCOM.2018.8599826}}


\bibitem[Karkehabadi et~al\mbox{.}(2024)]%
        {karkehabadi2024hlgm}
\bibfield{author}{\bibinfo{person}{Ali Karkehabadi}, \bibinfo{person}{Banafsheh~Saber Latibari}, \bibinfo{person}{Houman Homayoun}, {and} \bibinfo{person}{Avesta Sasan}.} \bibinfo{year}{2024}\natexlab{}.
\newblock \showarticletitle{HLGM: A novel methodology for improving model accuracy using saliency-guided high and low gradient masking}. In \bibinfo{booktitle}{\emph{2024 14th International Conference on Information Science and Technology (ICIST)}}. IEEE, \bibinfo{pages}{909--917}.
\newblock


\bibitem[Kuang et~al\mbox{.}(2025)]%
        {kuang2025ntndet}
\bibfield{author}{\bibinfo{person}{Shijie Kuang}, \bibinfo{person}{Zhe Quan}, \bibinfo{person}{Guoqi Xie}, \bibinfo{person}{Xiaomin Cai}, \bibinfo{person}{Xiaoqian Chen}, {and} \bibinfo{person}{Keqin Li}.} \bibinfo{year}{2025}\natexlab{}.
\newblock \showarticletitle{NtNDet: Hardware Trojan detection based on pre-trained language models}.
\newblock \bibinfo{journal}{\emph{Expert Systems with Applications}} (\bibinfo{year}{2025}), \bibinfo{pages}{126666}.
\newblock


\bibitem[Latibari et~al\mbox{.}(2025)]%
        {latibari2025optimizing}
\bibfield{author}{\bibinfo{person}{Banafsheh~Saber Latibari}, \bibinfo{person}{Houman Homayoun}, {and} \bibinfo{person}{Avesta Sasan}.} \bibinfo{year}{2025}\natexlab{}.
\newblock \showarticletitle{Optimizing Vision Transformers: Unveiling’Focus and Forget’for Enhanced Computational Efficiency}.
\newblock \bibinfo{journal}{\emph{IEEE Access}} (\bibinfo{year}{2025}).
\newblock


\bibitem[Latibari et~al\mbox{.}(2024)]%
        {latibari2024transformers}
\bibfield{author}{\bibinfo{person}{Banafsheh~Saber Latibari}, \bibinfo{person}{Najmeh Nazari}, \bibinfo{person}{Muhtasim~Alam Chowdhury}, \bibinfo{person}{Kevin~Immanuel Gubbi}, \bibinfo{person}{Chongzhou Fang}, \bibinfo{person}{Sujan Ghimire}, \bibinfo{person}{Elahe Hosseini}, \bibinfo{person}{Hossein Sayadi}, \bibinfo{person}{Houman Homayoun}, \bibinfo{person}{Soheil Salehi}, {et~al\mbox{.}}} \bibinfo{year}{2024}\natexlab{}.
\newblock \showarticletitle{Transformers: A Security Perspective}.
\newblock \bibinfo{journal}{\emph{IEEE Access}} (\bibinfo{year}{2024}).
\newblock


\bibitem[Lee et~al\mbox{.}(2021)]%
        {9446687}
\bibfield{author}{\bibinfo{person}{Woongsup Lee}, \bibinfo{person}{Seon~Yeob Baek}, {and} \bibinfo{person}{Seong~Hwan Kim}.} \bibinfo{year}{2021}\natexlab{}.
\newblock \showarticletitle{Deep-Learning-Aided RF Fingerprinting for NFC Security}.
\newblock \bibinfo{journal}{\emph{IEEE Communications Magazine}} \bibinfo{volume}{59}, \bibinfo{number}{5} (\bibinfo{year}{2021}), \bibinfo{pages}{96--101}.
\newblock
\href{https://doi.org/10.1109/MCOM.001.2000912}{doi:\nolinkurl{10.1109/MCOM.001.2000912}}


\bibitem[Li et~al\mbox{.}(2023)]%
        {li2023htrans}
\bibfield{author}{\bibinfo{person}{Yilin Li}, \bibinfo{person}{Shan Li}, {and} \bibinfo{person}{Haihua Shen}.} \bibinfo{year}{2023}\natexlab{}.
\newblock \showarticletitle{Htrans: Transformer-based method for hardware trojan detection and localization}. In \bibinfo{booktitle}{\emph{2023 IEEE 32nd Asian Test Symposium (ATS)}}. IEEE, \bibinfo{pages}{1--6}.
\newblock


\bibitem[Lin et~al\mbox{.}(2023)]%
        {lin2023hw}
\bibfield{author}{\bibinfo{person}{Yu-Zheng Lin}, \bibinfo{person}{Muntasir Mamun}, \bibinfo{person}{Muhtasim~Alam Chowdhury}, \bibinfo{person}{Shuyu Cai}, \bibinfo{person}{Mingyu Zhu}, \bibinfo{person}{Banafsheh~Saber Latibari}, \bibinfo{person}{Kevin~Immanuel Gubbi}, \bibinfo{person}{Najmeh~Nazari Bavarsad}, \bibinfo{person}{Arjun Caputo}, \bibinfo{person}{Avesta Sasan}, {et~al\mbox{.}}} \bibinfo{year}{2023}\natexlab{}.
\newblock \showarticletitle{Hw-v2w-map: Hardware vulnerability to weakness mapping framework for root cause analysis with gpt-assisted mitigation suggestion}.
\newblock \bibinfo{journal}{\emph{arXiv preprint arXiv:2312.13530}} (\bibinfo{year}{2023}).
\newblock


\bibitem[Liu et~al\mbox{.}(2024)]%
        {10643476}
\bibfield{author}{\bibinfo{person}{Fei Liu}, \bibinfo{person}{Heng Wang}, \bibinfo{person}{Pingfa Feng}, {and} \bibinfo{person}{Long Zeng}.} \bibinfo{year}{2024}\natexlab{}.
\newblock \showarticletitle{Integrated Circuit Packaging Defect Analysis and Deep Learning Detection Method}.
\newblock \bibinfo{journal}{\emph{IEEE Transactions on Components, Packaging and Manufacturing Technology}} \bibinfo{volume}{14}, \bibinfo{number}{9} (\bibinfo{year}{2024}), \bibinfo{pages}{1707--1719}.
\newblock
\href{https://doi.org/10.1109/TCPMT.2024.3447040}{doi:\nolinkurl{10.1109/TCPMT.2024.3447040}}


\bibitem[Makrani et~al\mbox{.}(2021)]%
        {makrani2021security}
\bibfield{author}{\bibinfo{person}{Hosein~Mohammadi Makrani} {et~al\mbox{.}}} \bibinfo{year}{2021}\natexlab{}.
\newblock \showarticletitle{Security threats in cloud rooted from machine learning-based resource provisioning systems}. In \bibinfo{booktitle}{\emph{Silicon Valley Cybersecurity Conference}}. Springer, \bibinfo{pages}{22--32}.
\newblock


\bibitem[Marali et~al\mbox{.}(2024)]%
        {marali2024hybrid}
\bibfield{author}{\bibinfo{person}{Mounesh Marali} {et~al\mbox{.}}} \bibinfo{year}{2024}\natexlab{}.
\newblock \showarticletitle{A hybrid transformer-based BERT and LSTM approach for vulnerability classification problems}.
\newblock \bibinfo{journal}{\emph{International Journal of Mathematics in Operational Research}} \bibinfo{volume}{28}, \bibinfo{number}{3} (\bibinfo{year}{2024}), \bibinfo{pages}{275--295}.
\newblock


\bibitem[Mashnoor et~al\mbox{.}(2025)]%
        {mashnoor2025llm}
\bibfield{author}{\bibinfo{person}{Nowfel Mashnoor}, \bibinfo{person}{Mohammad Akyash}, \bibinfo{person}{Hadi Kamali}, {and} \bibinfo{person}{Kimia Azar}.} \bibinfo{year}{2025}\natexlab{}.
\newblock \showarticletitle{LLM-IFT: LLM-Powered Information Flow Tracking for Secure Hardware}.
\newblock \bibinfo{journal}{\emph{arXiv preprint arXiv:2504.07015}} (\bibinfo{year}{2025}).
\newblock


\bibitem[Mirzaeian et~al\mbox{.}(2022)]%
        {mirzaeian2022adaptive}
\bibfield{author}{\bibinfo{person}{Ali Mirzaeian}, \bibinfo{person}{Zhi Tian}, \bibinfo{person}{Sai~Manoj PD}, \bibinfo{person}{Banafsheh~S Latibari}, \bibinfo{person}{Ioannis Savidis}, \bibinfo{person}{Houman Homayoun}, {and} \bibinfo{person}{Avesta Sasan}.} \bibinfo{year}{2022}\natexlab{}.
\newblock \showarticletitle{Adaptive-Gravity: A Defense Against Adversarial Samples}. In \bibinfo{booktitle}{\emph{2022 23rd International Symposium on Quality Electronic Design (ISQED)}}. IEEE, \bibinfo{pages}{96--101}.
\newblock


\bibitem[Mohamadsalehi(2022)]%
        {mohamadsalehi2022expansion}
\bibfield{author}{\bibinfo{person}{Mohamad Mohamadsalehi}.} \bibinfo{year}{2022}\natexlab{}.
\newblock \bibinfo{booktitle}{\emph{Expansion of Conforming to Interface Structured Adaptive Mesh Refinement Algorithm to Higher Order Elements and Crack Propagation}}.
\newblock \bibinfo{publisher}{The Ohio State University}.
\newblock


\bibitem[Mohmadsalehi and Soghrati(2022)]%
        {mohmadsalehi2022automated}
\bibfield{author}{\bibinfo{person}{Mohamad Mohmadsalehi} {and} \bibinfo{person}{Soheil Soghrati}.} \bibinfo{year}{2022}\natexlab{}.
\newblock \showarticletitle{An automated mesh generation algorithm for simulating complex crack growth problems}.
\newblock \bibinfo{journal}{\emph{Computer Methods in Applied Mechanics and Engineering}}  \bibinfo{volume}{398} (\bibinfo{year}{2022}), \bibinfo{pages}{115015}.
\newblock


\bibitem[Nasr et~al\mbox{.}(2024)]%
        {nasr2024improving}
\bibfield{author}{\bibinfo{person}{Abdurrahman Nasr}, \bibinfo{person}{Khalil Mohamed}, \bibinfo{person}{Mohamed Zaki}, {et~al\mbox{.}}} \bibinfo{year}{2024}\natexlab{}.
\newblock \showarticletitle{Improving Hardware Trojan Detection with Transformer-Based Power Analysis}.
\newblock  (\bibinfo{year}{2024}).
\newblock


\bibitem[Natsos and Symeonidis(2025)]%
        {natsos2025transformer}
\bibfield{author}{\bibinfo{person}{Dimosthenis Natsos} {and} \bibinfo{person}{Andreas~L Symeonidis}.} \bibinfo{year}{2025}\natexlab{}.
\newblock \showarticletitle{Transformer-based malware detection using process resource utilization metrics}.
\newblock \bibinfo{journal}{\emph{Results in Engineering}} (\bibinfo{year}{2025}), \bibinfo{pages}{104250}.
\newblock


\bibitem[Nazari et~al\mbox{.}(2024a)]%
        {nazari2024llm}
\bibfield{author}{\bibinfo{person}{Najmeh Nazari} {et~al\mbox{.}}} \bibinfo{year}{2024}\natexlab{a}.
\newblock \showarticletitle{LLM-FIN: Large Language Models Fingerprinting Attack on Edge Devices}. In \bibinfo{booktitle}{\emph{2024 25th International Symposium on Quality Electronic Design (ISQED)}}. IEEE, \bibinfo{pages}{1--6}.
\newblock


\bibitem[Nazari et~al\mbox{.}(2024b)]%
        {nazari2024securing}
\bibfield{author}{\bibinfo{person}{Najmeh Nazari} {et~al\mbox{.}}} \bibinfo{year}{2024}\natexlab{b}.
\newblock \showarticletitle{Securing On-Chip Learning: Navigating Vulnerabilities and Potential Safeguards in Spiking Neural Network Architectures}. In \bibinfo{booktitle}{\emph{2024 IEEE International Symposium on Circuits and Systems (ISCAS)}}. IEEE, \bibinfo{pages}{1--5}.
\newblock


\bibitem[Nazari et~al\mbox{.}(2024c)]%
        {nazari2024specscope}
\bibfield{author}{\bibinfo{person}{Najmeh Nazari} {et~al\mbox{.}}} \bibinfo{year}{2024}\natexlab{c}.
\newblock \showarticletitle{SpecScope: Automating Discovery of Exploitable Spectre Gadgets on Black-Box Microarchitectures}. In \bibinfo{booktitle}{\emph{2024 Design, Automation \& Test in Europe Conference \& Exhibition (DATE)}}. IEEE, \bibinfo{pages}{1--6}.
\newblock


\bibitem[Nazari et~al\mbox{.}(2024d)]%
        {nazari2024architectural}
\bibfield{author}{\bibinfo{person}{Najmeh Nazari}, \bibinfo{person}{Chongzhou Fang}, \bibinfo{person}{Hosein Mohammadi~Makrani}, \bibinfo{person}{Behnam Omidi}, \bibinfo{person}{Mahdi Eslamimehr}, \bibinfo{person}{Setareh Rafatirad}, \bibinfo{person}{Avesta Sasan}, \bibinfo{person}{Hossein Sayadi}, \bibinfo{person}{Khaled~N Khasawneh}, {and} \bibinfo{person}{Houman Homayoun}.} \bibinfo{year}{2024}\natexlab{d}.
\newblock \showarticletitle{Architectural Whispers: Robust Machine Learning Models Fingerprinting via Frequency Throttling Side-Channels}. In \bibinfo{booktitle}{\emph{Proceedings of the 61st ACM/IEEE Design Automation Conference}}. \bibinfo{pages}{1--6}.
\newblock


\bibitem[Nazari et~al\mbox{.}(2024e)]%
        {nazari2024forget}
\bibfield{author}{\bibinfo{person}{Najmeh Nazari}, \bibinfo{person}{Hosein~Mohammadi Makrani}, \bibinfo{person}{Chongzhou Fang}, \bibinfo{person}{Hossein Sayadi}, \bibinfo{person}{Setareh Rafatirad}, \bibinfo{person}{Khaled~N Khasawneh}, {and} \bibinfo{person}{Houman Homayoun}.} \bibinfo{year}{2024}\natexlab{e}.
\newblock \showarticletitle{Forget and Rewire: Enhancing the Resilience of Transformer-based Models against Bit-Flip Attacks}. In \bibinfo{booktitle}{\emph{33rd USENIX Security Symposium}}. \bibinfo{pages}{1349--1366}.
\newblock


\bibitem[Pal et~al\mbox{.}(2023)]%
        {pal2023temporal}
\bibfield{author}{\bibinfo{person}{Preetam Pal}, \bibinfo{person}{Pratik Chattopadhyay}, {and} \bibinfo{person}{Mayank Swarnkar}.} \bibinfo{year}{2023}\natexlab{}.
\newblock \showarticletitle{Temporal feature aggregation with attention for insider threat detection from activity logs}.
\newblock \bibinfo{journal}{\emph{Expert Systems with Applications}}  \bibinfo{volume}{224} (\bibinfo{year}{2023}), \bibinfo{pages}{119925}.
\newblock


\bibitem[Pan et~al\mbox{.}(2022)]%
        {pan2022-explainHMD}
\bibfield{author}{\bibinfo{person}{Zhixin Pan}, \bibinfo{person}{Jennifer Sheldon}, {and} \bibinfo{person}{Prabhat Mishra}.} \bibinfo{year}{2022}\natexlab{}.
\newblock \showarticletitle{Hardware-assisted malware detection and localization using explainable machine learning}.
\newblock \bibinfo{journal}{\emph{IEEE Trans. Comput.}} \bibinfo{volume}{71}, \bibinfo{number}{12} (\bibinfo{year}{2022}), \bibinfo{pages}{3308--3321}.
\newblock


\bibitem[Parpart et~al\mbox{.}(2024)]%
        {parpart2024transformer}
\bibfield{author}{\bibinfo{person}{Gavin Parpart}, \bibinfo{person}{Jonathan~H Tu}, \bibinfo{person}{Bradley Clymer}, \bibinfo{person}{Jung Lee}, {and} \bibinfo{person}{Jasen Babcock}.} \bibinfo{year}{2024}\natexlab{}.
\newblock \showarticletitle{Transformer Masked Autoencoders for RF Device Fingerprinting}. In \bibinfo{booktitle}{\emph{MILCOM 2024-2024 IEEE Military Communications Conference (MILCOM)}}. IEEE, \bibinfo{pages}{859--862}.
\newblock


\bibitem[Ravi et~al\mbox{.}(2023)]%
        {ravi2023vit4mal}
\bibfield{author}{\bibinfo{person}{Akshara Ravi}, \bibinfo{person}{Vivek Chaturvedi}, {and} \bibinfo{person}{Muhammad Shafique}.} \bibinfo{year}{2023}\natexlab{}.
\newblock \showarticletitle{Vit4mal: Lightweight vision transformer for malware detection on edge devices}.
\newblock \bibinfo{journal}{\emph{ACM Transactions on Embedded Computing Systems}} \bibinfo{volume}{22}, \bibinfo{number}{5s} (\bibinfo{year}{2023}), \bibinfo{pages}{1--26}.
\newblock


\bibitem[Saber~Latibari et~al\mbox{.}(2024)]%
        {saber2024iret}
\bibfield{author}{\bibinfo{person}{Banafsheh Saber~Latibari}, \bibinfo{person}{Soheil Salehi}, \bibinfo{person}{Houman Homayoun}, {and} \bibinfo{person}{Avesta Sasan}.} \bibinfo{year}{2024}\natexlab{}.
\newblock \showarticletitle{IRET: Incremental Resolution Enhancing Transformer}. In \bibinfo{booktitle}{\emph{Proceedings of the Great Lakes Symposium on VLSI 2024}}. \bibinfo{pages}{620--625}.
\newblock


\bibitem[Sayadi et~al\mbox{.}(2022)]%
        {sayadi2022towards}
\bibfield{author}{\bibinfo{person}{Hossein Sayadi}, \bibinfo{person}{Mehrdad Aliasgari}, \bibinfo{person}{Furkan Aydin}, \bibinfo{person}{Seetal Potluri}, \bibinfo{person}{Aydin Aysu}, \bibinfo{person}{Jack Edmonds}, {and} \bibinfo{person}{Sara Tehranipoor}.} \bibinfo{year}{2022}\natexlab{}.
\newblock \showarticletitle{Towards ai-enabled hardware security: Challenges and opportunities}. In \bibinfo{booktitle}{\emph{2022 IEEE 28th International Symposium on On-Line Testing and Robust System Design (IOLTS)}}. IEEE, \bibinfo{pages}{1--10}.
\newblock


\bibitem[Sayadi et~al\mbox{.}(2020)]%
        {sayadi2020stealthminer}
\bibfield{author}{\bibinfo{person}{Hossein Sayadi}, \bibinfo{person}{Yifeng Gao}, \bibinfo{person}{Hosein Mohammadi~Makrani}, \bibinfo{person}{Tinoosh Mohsenin}, \bibinfo{person}{Avesta Sasan}, \bibinfo{person}{Setareh Rafatirad}, \bibinfo{person}{Jessica Lin}, {and} \bibinfo{person}{Houman Homayoun}.} \bibinfo{year}{2020}\natexlab{}.
\newblock \showarticletitle{Stealthminer: Specialized time series machine learning for run-time stealthy malware detection based on microarchitectural features}. In \bibinfo{booktitle}{\emph{Proceedings of the 2020 on Great Lakes Symposium on VLSI}}. \bibinfo{pages}{175--180}.
\newblock


\bibitem[Sayadi et~al\mbox{.}(2024)]%
        {sayadi-ISQED2024-survey}
\bibfield{author}{\bibinfo{person}{Hossein Sayadi}, \bibinfo{person}{Zhangying He}, \bibinfo{person}{Hosein~Mohammadi Makrani}, {and} \bibinfo{person}{Houman Homayoun}.} \bibinfo{year}{2024}\natexlab{}.
\newblock \showarticletitle{Intelligent malware detection based on hardware performance counters: A comprehensive survey}. In \bibinfo{booktitle}{\emph{2024 25th International Symposium on Quality Electronic Design (ISQED)}}. IEEE, \bibinfo{pages}{1--10}.
\newblock


\bibitem[Sayadi et~al\mbox{.}(2019)]%
        {sayadi-DATE2019-2smart}
\bibfield{author}{\bibinfo{person}{Hossein Sayadi}, \bibinfo{person}{Hosein~Mohammadi Makrani}, \bibinfo{person}{Sai Manoj~Pudukotai Dinakarrao}, \bibinfo{person}{Tinoosh Mohsenin}, \bibinfo{person}{Avesta Sasan}, \bibinfo{person}{Setareh Rafatirad}, {and} \bibinfo{person}{Houman Homayoun}.} \bibinfo{year}{2019}\natexlab{}.
\newblock \showarticletitle{2smart: A two-stage machine learning-based approach for run-time specialized hardware-assisted malware detection}. In \bibinfo{booktitle}{\emph{2019 Design, Automation \& Test in Europe Conference \& Exhibition (DATE)}}. IEEE, \bibinfo{pages}{728--733}.
\newblock


\bibitem[Sayadi et~al\mbox{.}(2018)]%
        {sayadi-ensemble-DAC18}
\bibfield{author}{\bibinfo{person}{Hossein Sayadi}, \bibinfo{person}{Nisarg Patel}, \bibinfo{person}{Avesta Sasan}, \bibinfo{person}{Setareh Rafatirad}, {and} \bibinfo{person}{Houman Homayoun}.} \bibinfo{year}{2018}\natexlab{}.
\newblock \showarticletitle{Ensemble learning for effective run-time hardware-based malware detection: A comprehensive analysis and classification}. In \bibinfo{booktitle}{\emph{Proceedings of the 55th Annual Design Automation Conference}}. \bibinfo{pages}{1--6}.
\newblock


\bibitem[Seneviratne et~al\mbox{.}(2022)]%
        {seneviratne2022self}
\bibfield{author}{\bibinfo{person}{Sachith Seneviratne}, \bibinfo{person}{Ridwan Shariffdeen}, \bibinfo{person}{Sanka Rasnayaka}, {and} \bibinfo{person}{Nuran Kasthuriarachchi}.} \bibinfo{year}{2022}\natexlab{}.
\newblock \showarticletitle{Self-supervised vision transformers for malware detection}.
\newblock \bibinfo{journal}{\emph{IEEE Access}}  \bibinfo{volume}{10} (\bibinfo{year}{2022}), \bibinfo{pages}{103121--103135}.
\newblock


\bibitem[Shen et~al\mbox{.}(2021)]%
        {shen2021radio}
\bibfield{author}{\bibinfo{person}{Guanxiong Shen}, \bibinfo{person}{Junqing Zhang}, \bibinfo{person}{Alan Marshall}, \bibinfo{person}{Mikko Valkama}, {and} \bibinfo{person}{Joseph Cavallaro}.} \bibinfo{year}{2021}\natexlab{}.
\newblock \showarticletitle{Radio frequency fingerprint identification for security in low-cost IoT devices}. In \bibinfo{booktitle}{\emph{2021 55th Asilomar conference on signals, systems, and computers}}. IEEE, \bibinfo{pages}{309--313}.
\newblock


\bibitem[Shrivastwa et~al\mbox{.}(2021)]%
        {shrivastwa2021multi}
\bibfield{author}{\bibinfo{person}{Ritu-Ranjan Shrivastwa}, \bibinfo{person}{Sylvain Guilley}, {and} \bibinfo{person}{Jean-Luc Danger}.} \bibinfo{year}{2021}\natexlab{}.
\newblock \showarticletitle{Multi-source fault injection detection using machine learning and sensor fusion}. In \bibinfo{booktitle}{\emph{International Conference on Security and Privacy}}. Springer, \bibinfo{pages}{93--107}.
\newblock


\bibitem[Srivastava et~al\mbox{.}(2024)]%
        {srivastava2024scar}
\bibfield{author}{\bibinfo{person}{Amisha Srivastava} {et~al\mbox{.}}} \bibinfo{year}{2024}\natexlab{}.
\newblock \showarticletitle{SCAR: Power Side-Channel Analysis at RTL Level}.
\newblock \bibinfo{journal}{\emph{IEEE Transactions on Very Large Scale Integration (VLSI) Systems}} (\bibinfo{year}{2024}).
\newblock


\bibitem[Vaswani et~al\mbox{.}(2017)]%
        {vaswani2017attention}
\bibfield{author}{\bibinfo{person}{Ashish Vaswani} {et~al\mbox{.}}} \bibinfo{year}{2017}\natexlab{}.
\newblock \showarticletitle{Attention is all you need}.
\newblock \bibinfo{journal}{\emph{Advances in neural information processing systems}}  \bibinfo{volume}{30} (\bibinfo{year}{2017}).
\newblock


\bibitem[Wang et~al\mbox{.}(2021)]%
        {wang2021enabling}
\bibfield{author}{\bibinfo{person}{Han Wang}, \bibinfo{person}{Hossein Sayadi}, \bibinfo{person}{Sai Manoj~Pudukotai Dinakarrao}, \bibinfo{person}{Avesta Sasan}, \bibinfo{person}{Setareh Rafatirad}, {and} \bibinfo{person}{Houman Homayoun}.} \bibinfo{year}{2021}\natexlab{}.
\newblock \showarticletitle{Enabling micro AI for securing edge devices at hardware level}.
\newblock \bibinfo{journal}{\emph{IEEE Journal on Emerging and Selected Topics in Circuits and Systems}} \bibinfo{volume}{11}, \bibinfo{number}{4} (\bibinfo{year}{2021}), \bibinfo{pages}{803--815}.
\newblock


\bibitem[Wang et~al\mbox{.}(2024)]%
        {wang2024deep}
\bibfield{author}{\bibinfo{person}{Xiaobin Wang}, \bibinfo{person}{Shuang Gao}, \bibinfo{person}{Jianlan Guo}, \bibinfo{person}{Chu Wang}, \bibinfo{person}{Liping Xiong}, {and} \bibinfo{person}{Yuntao Zou}.} \bibinfo{year}{2024}\natexlab{}.
\newblock \showarticletitle{Deep learning-based integrated circuit surface defect detection: Addressing information density imbalance for industrial application}.
\newblock \bibinfo{journal}{\emph{International Journal of Computational Intelligence Systems}} \bibinfo{volume}{17}, \bibinfo{number}{1} (\bibinfo{year}{2024}), \bibinfo{pages}{29}.
\newblock


\bibitem[Wu et~al\mbox{.}(2018)]%
        {wu2018deep}
\bibfield{author}{\bibinfo{person}{Qingyang Wu}, \bibinfo{person}{Carlos Feres}, \bibinfo{person}{Daniel Kuzmenko}, \bibinfo{person}{Ding Zhi}, \bibinfo{person}{Zhou Yu}, \bibinfo{person}{Xin Liu}, {and} \bibinfo{person}{Xiaoguang ‘Leo’Liu}.} \bibinfo{year}{2018}\natexlab{}.
\newblock \showarticletitle{Deep learning based RF fingerprinting for device identification and wireless security}.
\newblock \bibinfo{journal}{\emph{Electronics Letters}} \bibinfo{volume}{54}, \bibinfo{number}{24} (\bibinfo{year}{2018}), \bibinfo{pages}{1405--1407}.
\newblock


\bibitem[Ye et~al\mbox{.}(2024)]%
        {YE2024103971}
\bibfield{author}{\bibinfo{person}{Junjian Ye} {et~al\mbox{.}}} \bibinfo{year}{2024}\natexlab{}.
\newblock \showarticletitle{Detecting command injection vulnerabilities in Linux-based embedded firmware with LLM-based taint analysis of library functions}.
\newblock \bibinfo{journal}{\emph{Computers \& Security}}  \bibinfo{volume}{144} (\bibinfo{year}{2024}), \bibinfo{pages}{103971}.
\newblock


\bibitem[Yu et~al\mbox{.}(2021)]%
        {9631481}
\bibfield{author}{\bibinfo{person}{Yingchao Yu}, \bibinfo{person}{Shuitao Gan}, {and} \bibinfo{person}{Xiaojun Qin}.} \bibinfo{year}{2021}\natexlab{}.
\newblock \showarticletitle{firm VulSeeker: BERT and Siamese based Vulnerability for Embedded Device Firmware Images}. In \bibinfo{booktitle}{\emph{2021 IEEE Symposium on Computers and Communications (ISCC)}}. \bibinfo{pages}{1--7}.
\newblock


\end{thebibliography}


\end{document}